\documentclass[manuscript,onecolumn]{geophysics}
\usepackage{graphicx}
\usepackage{subfigure}
\usepackage[english]{babel}
\usepackage{amsmath, amsthm, amssymb}
\usepackage{mathtools}
\usepackage{geometry}
\usepackage{setspace}
\usepackage[dvipsnames]{xcolor}
\usepackage{rotating}
\newcommand{\x}{{ \bf{x}}}
\begin{document}
\title{Identifying the Unique Geochemical Fingerprints of
 Omo Group Beds around Lake Turkana, Kenya by Modified Gaussian Discrimination Analysis}

\renewcommand{\thefootnote}{\fnsymbol{footnote}}
%\ms{GEO-2021-0306.R2} % manuscript number
\address{
\footnotemark[1]Department of Geology and Geophysics, University of Utah \\
\footnotemark[2]University of North Carolina at Chapel Hill\\
\footnotemark[3]Viridien \\
}
\author{Gerard T. Schuster\footnotemark[1] and Shihang Feng\footnotemark[2]\footnotemark[3]}

\footer{Manuscript}
\lefthead{Schuster&Gathogo}
\righthead{Geochemical Fingerprints}

\maketitle
\begin{abstract}
A modified Gaussian Discriminant Analysis (GDA) is used with an optimal search strategy
to identify the unique geochemical fingerprints
of six different geological beds in the Lake Turkana area.
Three-hundred samples were collected from six different beds in the Omo Group of Lake Turkana,
where each sample consisted of the PPMs of 11 different chemical compound.
The GDA analysis discovered a unique combination of three tuff compounds that
can uniquely identify one
of the five beds, where a 6th bed is excluded because only 7 samples were
collected from it.
These geochemical fingerprints are important because
any Turkana hominin fossil in the Omo Group can now be efficiently identified
by  matching the geochemical fingerprint  of a fossil's host rock to the fingerprint of the bed in which it is found.
In general, the GDA search strategy can be a powerful tool for efficiently identifying
unique
class fingerprints of
high-dimensional data.

\end{abstract}

\section{Introduction}

The Lake Turkana region in Kenya is a well-known site for many hominin fossils
(Gathogo and Brown, 2006) that date back to about 4.3 million years ago (Myr). These fossils have provided crucial information for understanding the evolution of hominins during this period.

Determining the age of these fossils is done by assigning their age to that of the bed they were embedded. The age of the formation can be determined through several methods, including  isotopic dating and geochemical correlation of volcanic materials, which rely on the presence of volcanic ash or rocks in the formation.

\begin{figure}
\includegraphics[width=5.8in,height=2.5in]{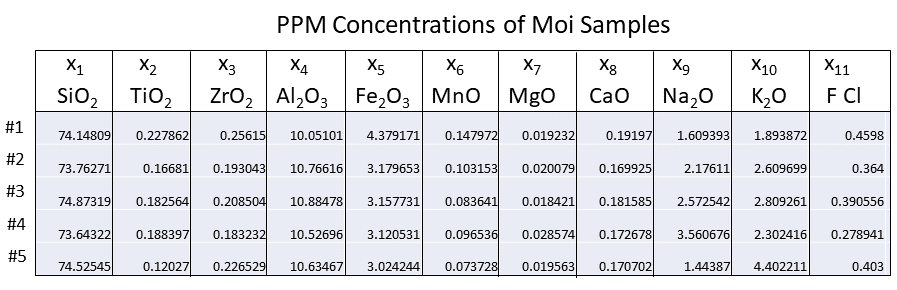}
\caption{Table of ppms for geochemical compounds in 5 samples
of rocks from the Moi formation in Turkana. The top row of labels denotes the type
of compound and the next five rows depict the ppm content
in these samples.
The coordinate labels for the compounds in the top row are, from left to right,
$x_1, x_2,...,x_{11}$. There are a total of 300 samples from 6 different formations.}
\label{Screen.Data.Summary}
\end{figure}

However, identifying the tuffs associated with a hominin discovery is not always easy, and geochemical analysis is required. Ideally, the geochemical signature of a tuff should be unique, allowing it to be distinguished from tuffs in the other formations.
For the Turkana example, the ppms $x_i$ of eleven different compounds, i.e. $i\in \{1,2,...,11\}$, in  Figure~\ref{Screen.Data.Summary}
were measured by an electron microprobe for each tuff sample. A total of three hundred samples were obtained from six different beds, and the ppms for each sample is described by an $11 \times 1$ vector
 $(x_1,x_2,...,x_{11})$ in an eleven-dimensional space. It is hoped that the points from any one formation form a point cluster
will be distinct from the other five point clusters. In this way, a new rock sample
can be analyzed and uniquely matched to the formation and its associated age.

\begin{figure}
\includegraphics[width=5.8in]{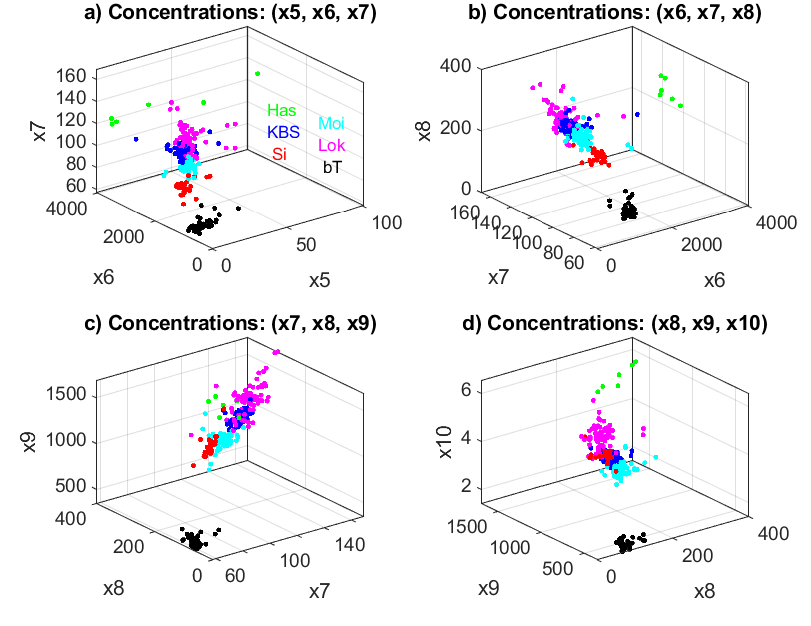}
  \caption {Three-dimensional plots of 300  Turkana geochem points $(x_{i}, x_{j}, x_k)$, where each color corresponds to a different formation name abbreviated in a).
  There is too much overlap in the colored point clouds to
  assign a unique geochemical fingerprint of three compounds to
  all six beds.
  The units along axes are parts per million (ppm) and each coordinate $x_i$ corresponds to
  one of the compounds in Figure~\ref{Screen.Data.Summary}.}
  \label{3D.Concentration.B}
\end{figure}

Unfortunately, these point clusters most often overlap one another for different tuffs,
as shown in Figure~\ref{3D.Concentration.B} for different
3D subspaces.
For these examples, there is too much overlap of the point
clouds to allow for a unique assignment of a sample to any one formation.
Visually searching for non-overlapping point clouds in all triplet
combinations of $x_i-x_j-x_k$ for $i,j,k \in \{1,2,...,11\}$ is
too tedious.

\begin{figure}
\includegraphics[width=5.8in]{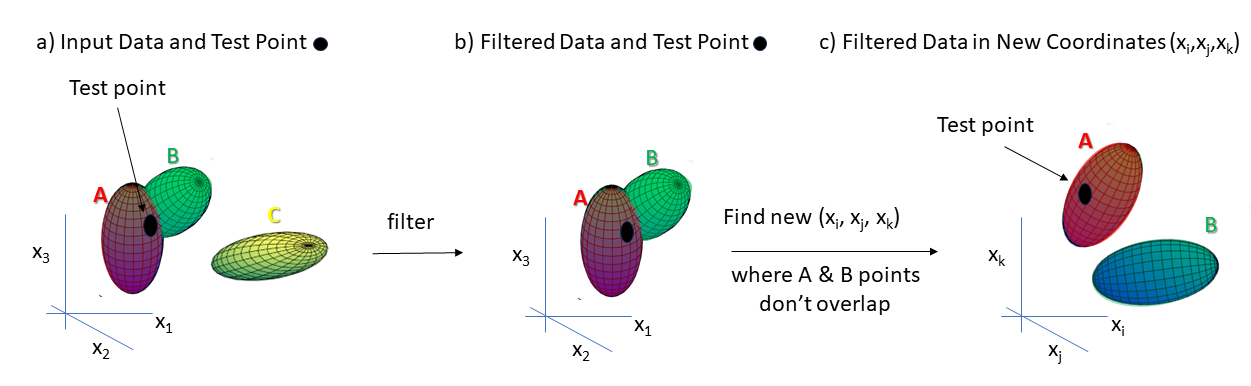}
  \caption {Schematic for determining the membership of
  a test point when it belongs to two
  overlapping ellipsoids. a) Black test point belongs to both
  the red and green ellipsoids in the coordinate
  system $(x_1,x_2,x_3)$. All points outside the red and
green ellipsoids are eliminated to give b). c) A new coordinate system $(x_i,x_j,x_k)$ is
found where the $A$ and $B$ ellipsoids
 are separable, and the test point is now identified in c)
 as belonging to formation A.}
  \label{2ptOverlap}
\end{figure}

We propose using a modified Gaussian Discriminant Analysis (Bishop, 2006)  to fit
each point cloud to its Gaussian distribution. After fitting,
we can
then efficiently identify the triplet of coordinates where
the Gaussians
don't overlap with one another within two standard deviations.
If there are no such triplets, then
the triplets of coordinates are sought where there are only two overlapping Gaussians. In this case,
these two overlapping Gaussians for beds A and B in Figure~\ref{2ptOverlap}a
are further tested by filtering out all points except for those in
formations A and B to get Figure~\ref{2ptOverlap}b, Then, we find a  unique combination of triplet coordinates $(x_i,x_j,x_k)$ where the point clouds in
beds A and B are non-overlapping. For example, a test point in the coordinate
system $(x_1,x_2,x_3)$ is found that belongs in the ellipsoids
of both A and B, but not other ellipsoids. After muting out all the points except those in A and B, and then transforming the A and B points to the new coordinate system $(x_i, x_j, x_k)$ indicates
which formation the test point belongs to.
This is similar to panning for gold, where screens with finer mesh sizes are sequentially used
to sift out all but the finest gold nuggets.
In our case, we are panning for the unique formation fingerprint of a sample.

This paper is divided into 3 sections. The first one is the introduction, which is
followed by a description of the traditional methods for signature identification
such as Principal Component Analysis (PCA) for
reducing the dimensionality of large-dimensional data sets.
The next section describes how modified Gaussian Discriminant Analysis (GDA) can be used
to uniquely identify
the geochemical fingerprints of different formations.
Results are presented for the geochemical data obtained from six
different formations in the Lake Turkana area that are important sites
of hominin fossils.
Finally, the summary section is included.

\section{Formation Fingerprint  by Principal Component Analysis}

Identifying a subspace where the point clouds of different formations do not overlap one another can be very difficult. For example, plotting the 11-dimensional
Turkana samples in a three-dimensional coordinate system $(x_i, x_j, x_k)$ results in the
point-cluster clouds in Figure~\ref{3D.Concentration.B}a. Here, there is significant overlap of point clouds with one another.
Other
combinations of  $1\leq i, j, k \leq 11$ shown in Figure~\ref{3D.Concentration.B}b-\ref{3D.Concentration.B}d show a similar overlap of point clouds. For these examples, there is too much overlap of the point
clouds to allow for a unique assignment of a sample to any one formation.

We can visually
 inspect all\footnote{How many different random samples of size 3 can be obtained from a population whose size is 11? The answer (DeGroot, 1975) is known as the combinatoric
$\binom{N}{k} ~=~ \frac{N!}{k!(N-k)!} $.} $\binom{N}{k} ~=~ \binom{11}{3}~= ~ 165$
3D plots for different triplet combinations of  $(x_{i}, x_{j}, x_k)$, but this is far too tedious to perform in practice. It's even worse if, for example, there are 30 compounds so that there are $\binom{30}{3} ~= ~ 4060$ 3D graphs to inspect for separability of the point clouds.
Can we reduce the dimensionality of such data so that the point clouds can be separable?
Principal component analysis is one tool that can reduce the dimensionality
of high-dimensional point clusters (Bishop, 2006) and perhaps achieve separability of the point clouds
for each formation.

\subsection{Principal Component Analysis}

Principal component analysis (PCA) is a statistical method used for dimensionality reduction in data analysis (Bishop, 2006). After demeaning the data,
PCA transforms it into  a new coordinate system where the first coordinate (principal component) accounts for the largest variance in the data, the second coordinate accounts for the second-largest variance, and so on. Thus,  PCA method
can be useful in reducing the number of variables needed to describe a dataset, while retaining the most important information.

\begin{figure}
\includegraphics[width=5.8in]{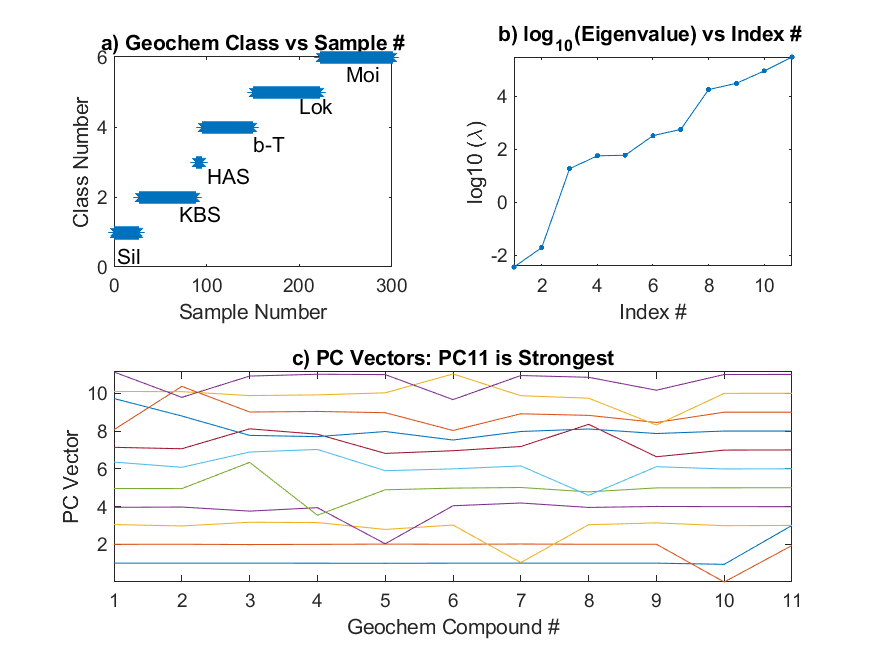}
  \caption {a) 300 data points plotting
  formation class against sample number. Each data point is represented
   by an $11 \times 1$ vector ${\bf x} =(x_1,x_2,...,x_{11})$, and each demeaned data vector
  forms a row in the $300 \times 11$ matrix $\bf X$.
  The eigenvalues and eigenvectors of the covariance matrix ${\bf X}^T{\bf X}$ are respectively displayed in  b) and c).}
    \label{Data.Summary}
\end{figure}

To apply PCA analysis, we demean the 300 $11 \times 1$ data vectors and assemble them into the $300 \times 11 $ data matrix denoted by $\bf X$ . The covariance matrix $\bf C$  is formed as ${\bf C} = {\bf X}^T {\bf X}$, and its eigenvalues and eigenvectors are depicted in Figure~\ref{Data.Summary}.
In this way we might be able to achieve separability of the point clouds
for each formation.

\begin{figure}

\includegraphics[width=5.8in]{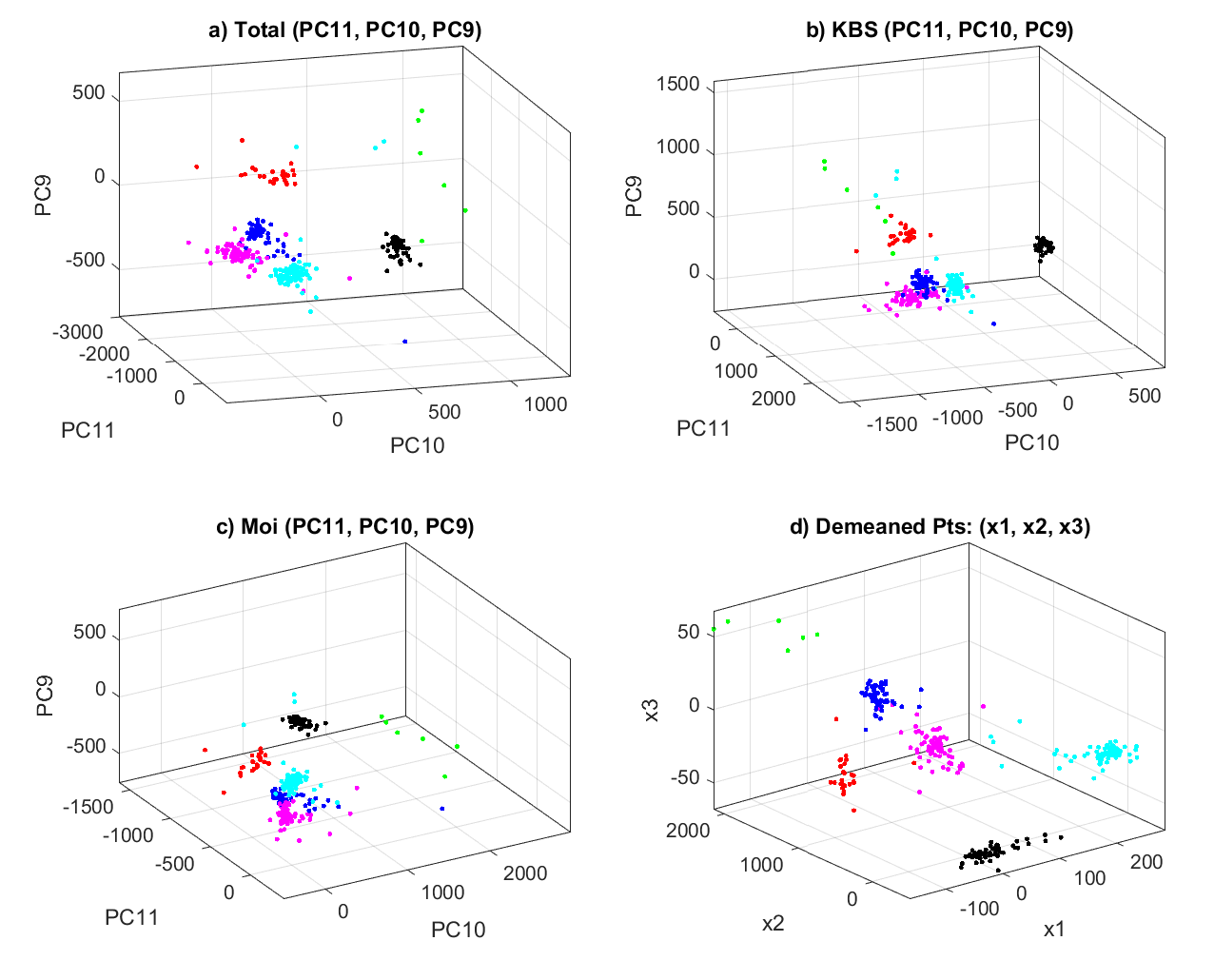}
  \caption {Demeaned data points projected onto the a) total
  Principal Component axes, b) KBS Principal Component axes,
  and c) Moiti (abbreviated as Moi)  Principal Component axes.
  Using the GDA search strategy, the
demeaned data plotted w/r to the $x_1-x_2-x_3$ axes
is in d), which is the best choice of axes where the point clouds are well separated
from one another.}
    \label{Data.Summary1}
\end{figure}

The 300 data points projected onto the principal axes
as
\begin{eqnarray}
\x'&=&({\bf PC11}\cdot \x,~ {\bf PC10}\cdot \x,~ {\bf PC9}\cdot \x),
\end{eqnarray}
are shown in Figure~\ref{Data.Summary1}a, where there is more of a separation of the six classes of point clouds compared to the point clouds in Figure~\ref{3D.Concentration.B}.
Here, the unit vectors along the PC axes are denoted as ${\bf PC11}$,
${\bf PC10}$, and ${\bf PC9}$. The largest eigenvalue is $\lambda_{11}$, the next largest is $\lambda_{10}$,
and $\lambda_{9}$ is the third largest, and are associated with the
unit vectors along the PC axes. Similar to Figure~\ref{Screen.Data.Summary}b,
the eigenvalues become smaller by almost an order-of-magnitude with decreasing subscript number. We denote these PC axes as {\it Total PC } axes because they are formed from the
300 data vectors from all six formations.

Another approach is to form a covariance matrix from the $11 \times 1$ data vectors that are only sampled from a
 single formation, and project the data points from all the formations
 onto these first three principal component vectors. For example, the strongest PC vectors for the {\it KBS} formation are denoted as ${\bf PC11}^{KBS}, ~{\bf  PC10}^{KBS}, $ and$ ~{\bf PC9}^{KBS}$, and the 3D coordinates of any $11 \times 1$ data vector $\bf x$ are obtained as
\begin{eqnarray}
{\bf x}'  &=&( {\bf PC11}^{KBS} \cdot {\bf x}, ~{\bf PC10}^{KBS} \cdot {\bf x}, ~{\bf PC9}^{KBS} \cdot {\bf x}),
\end{eqnarray}
where $\bf x'$ represents the vector $\bf x$ projected onto the
${\bf PC11}^{KBS}, {\bf  PC10}^{KBS}, $ and $ {\bf PC9}^{KBS}$ axes.
Figure~\ref{Data.Summary1}b depicts the results of projecting
300 data points onto the strongest three PC axes of the $KBS$ formation,
 and shows less point-cloud separability compared to Figure~\ref{Data.Summary1}a.
 Similarly, Figure~\ref{Data.Summary1}c shows the data points projected onto the
 PC axes from the KBS data and show less cloud separability than that shown in Figure~\ref{Data.Summary1}a.

The question now arises: is there a more effective way to explore a high-dimensional space
for separable point clouds? The next section answers this question by using
a systematic search method combined with Gaussian Discriminant Analysis (GDA).
As an example,  GDA analysis discovered the $(x_1,x_2,x_3)$ coordinate
system in
Figure~\ref{Data.Summary1}d which maximizes the separation of different point clouds.

\section{Formation Fingerprint  by Modified Gaussian Discriminant Analysis}

Gaussian Discriminant Analysis fits a $D$-dimensional Gaussian distribution
\begin{eqnarray}
p(\x|c,{\boldsymbol \Sigma}_c,{\boldsymbol \mu}_c )&=&
\frac{1}{\sqrt{(2\pi )^D |{\boldsymbol \Sigma}_c|}}e^{-\frac{1}{2}(\x-{\boldsymbol \mu}_c)^T{\boldsymbol \Sigma}_c^{-1}
(\x-{\boldsymbol \mu}_c)},
\label{GDA.eq5}
\end{eqnarray}
to
clouds of data points $\x$ that live in an $D-$dimensional space (Bishop, 2006). Here,
${\boldsymbol \Sigma}_c$ is the $11 \times 11$ covariance matrix for the
Turkana data associated with the class
$c$, and  ${\boldsymbol \mu}_c$ is the $11 \times 1$ mean vector for the points in class $c$.
For the Turkana example,
 $c \in \{1,2,...,6\}$. The 300 Turkana points can now be used
 to construct a Gaussian density
 distribution for each of the six formations.

The Gaussian decision volume $f(\x)_c$ is defined by saying that a point
$\x$ exclusively belongs to the class $c$ if
\begin{eqnarray}
f(\x)_c&=&p(c|\x)-\alpha \ge 0,
\label{Eq.overlap}
\end{eqnarray}
 for each of the separable cluster classes $c\in \{1,2,...,6\}$.
 Here,
\ $\alpha>0$ is  very small, say $\alpha=0.01$.
This means that, for a point $\x$ that belongs to class $c$, $\alpha$ can be chosen so
that more than 95\% of the points in cluster $c$ are within the decision volume
defined by $f(\x)_c$. Thus, if a new untested data point $\x$ satisfies
the above decision equation then it belongs to the class $c$ formation. This assumes that the
6 clusters don't have overlapping Gaussian distributions.

The workflow for GDA analysis of the Lake Turkana data is the following.
\begin{enumerate}
\item Choose a unique triplet of integer values  $(i,j,k) \in \{1,2,...,11\}$ and
$i\neq j \neq k$, where each coordinate $x_i$ is uniquely associated with one of the 11 chemical compounds listed in Figure~\ref{Screen.Data.Summary}.
This triplet of integers defines the 3D coordinate axes for $(x_i, x_j, x_k)$.

\item Compute the $1 \times 3$ mean vector  ${\boldsymbol \mu}_c$ and $3 \times 3$
covariance matrix
${\boldsymbol \Sigma}_c$
of each formation class $c \in \{1,2,...,6\}$ in the $(x_i, x_j, x_k)$ coordinates.
These quantities are then used
in equation~\ref{GDA.eq5} to compute the {\it generative} conditional probability function $p(\x|c,{\boldsymbol \Sigma}_c,{\boldsymbol \mu}_c )$.
We will denote this conditional function as $p(\x|c)$.

\item Compute the {\it discriminative} probability function $p(c|\x)$ by Bayes' Rule
(Bishop, 2006):
\begin{eqnarray}
p(c|\x) &=& \frac{p(\x|c) p(c)}{p(\x)},
\label{GDA.eq8}
\end{eqnarray}
where, for convenience, we assume $p(c)=1/6~~for ~c\in \{1,2,...,6\}$,
and $p(\x|c)$ is computed from equation~\ref{GDA.eq5}.
This allows us to compute $p(\x)=\sum_{c=1}^6 p(\x|c)p(c)$.

\item Identify the triplets of coordinates $(x_i,x_j,x_k)$ where the point clusters are separable
where  there is no pair-wise intersection of the six decision volumes, i.e.
\begin{eqnarray}
 f(\x)_{c'}{\displaystyle \cap } f(\x)_c = 0~~~\forall \x~~,~~\forall ~ c,c' \in \{1,2,...,6\} ~and~c\neq c'.
\end{eqnarray}
If the above cluster-separation condition is satisfied then $(i,j,k)$ is a triplet of indices
where the associated Gaussian ellipsoids for the rock formations do not overlap.

\item Repeat steps 1-4 until all possible triplet indices are tested.
If there are no triplets of indices where all the ellipsoids are separable, then
the indices can be searched where there are only two overlapping  ellipsoids.
In this case the {\it sieve} scheme outlined in Figure~\ref{Screen.Data.Summary}
can be implemented. This can be extended to indices where
only three clusters overlap.
A snippet from the MATLAB code for identifying ellipsoids without overlap
is below.
\begin{footnotesize}
\begin{verbatim}
%%%%%%%%%%%%%%%%%%%%%%%%%%%%%%%%%%%%%%%%%%%%%%%%%%%%%%%%%%%%%%%%%%%%%%%%%
% Sieve extracts the coordinate combinations of (ie1,ie2,ie3,ir) where all samples xx=xxx{ir}
% in formation ir are outside of Test ellipse. The ppms of the Test samples are denoted
% by the coordinates Test0=[Test(:,ie1) Test(:,ie2) Test(:,ie3)].
%
% lim     - input- threshold probability where a comparison sample is considered outside Test
%                  ellipsoid if p<lim. Here, 1x3 vector r=m-[xx(is,ie1) xx(is,ie2) xx(is,ie3)],
%                   p=al*exp(-0.5*r*XXI*r'), m is 1x3 mean vector of Test data and XXI is 3x3 inverse
%                  of Test covariance matrix, and vector xx contains ppms of comparison rock with
%                  sample number 'is'.
%
% Test0   - input- Nele x 3 vector of ppms for 3 compounds in Test0 rock, such as the Moi formation
%                  associated with 78x3 vector Test0=[Test(:,ie1) Test(:,ie2) Test(:,ie3)]. Here,
%                  (ie1,ie2,ie3) are integers 1-11 that indicate 11 types of compound, and we assume
%                  78 geochem  samples from the test0 formation for Moi, where Nele=78.
%
% xxx     - input- Geochem measurements xxx={[Sil] [KBS] [Has] [bT] [Lok] [Moi]}
%                  for the rock-formation matrices [Sil] [KBS] [Has] [bT] [Lok] [Moi].For
%                  example, [Moi] is a 78x11 matrix with ppms of 78 rock samples for each
%                  of the 11 measured compounds (e.g.,  FCl).
%
%tally{itally} -output- tally={ie1 ie2 ie3 ir Ntest Score(ie1,ie2,ie3,ir)}; Tally the information
%                       of rock formation samples that are outside the Test0 ellipse. The Test0 ellipse
%                   has the index Ntest that can take on any integer between 1 and 6.
%
%%%%%%%%%%%%%%%%%%%%%%%%%%%%%%%%%%%%%%%%%%%%%%%%%%%%%%%%%%%%%%%%%%%%%%%%%
load data;
[Nele N]=size(Test);Nrock=6;lim=0.01
for ie1=1:9           % Loop over integers ie1, ie2, ie3 of compound coordinates
 for ie2=ie1+1:10
    for ie3=ie2+1:11
% Loop over element coordinates in Test0=[Test(:,ie1) Test(:,ie2) Test(:,ie3)];
      Test0=[Test(:,ie1) Test(:,ie2) Test(:,ie3)];   % Nele x 3 Test Data Matrix
      m = mean(Test0);                               % 1x3 Test mean vector
      CC=cov(Test0);                                 % 3x3 Test  Covariance Matrix
      CC=cov(Test0)+eye(3)*.0001*max(abs(CC(:)));    % 3x3 Regularized Test Covariance matrix
      al=1;  itally=0;                               % Normalize Gaussian of Test Points
      XXI=inv(CC);
       for ir=1:Nrock
         xx=xxx{ir};                             % Nele x 3 matrix for the Comparison Rock
         [ii,pp]=search(ir,XXI,m,xx,ie1,ie2,ie3,al,lim); % Compute Probabilities p of Comparison Pts
                                                 % in xx being within p=lim of Test Gaussian
         [samp1, Nele1]= size(xxx{ir});
          Score(ie1,ie2,ie3,ir)=100*ii/samp1; % Score=100 if p<lim for ir Comparison rock formation
            if Score(ie1,ie2,ie3,ir)==100;
              itally=itally+1;
              tally{itally}=[ie1 ie2 ie3 ir Ntest Score(ie1,ie2,ie3,ir)]; % Tally information of
                                                       % point-cloud formations outside Test0 ellipsoid
            end
        end % If itally=5, then this triplet (ie1,ie2,ie3) of coordinates separates all the point clouds
     end
  end
end

\end{verbatim}
\end{footnotesize}
\end{enumerate}

Figure~\ref{Data.Summary1}d shows an
example of the Turkana points plotted in the non-overlapping
 set of point clouds in the $(x_1,x_2,x_3)$ axes.
 Here, the point clouds are mostly well separated, and even more so compared to those
 in the Principal Component axes in Figure~\ref{Data.Summary1}a-\ref{Data.Summary1}c.

\begin{figure}
\includegraphics[width=5.8in,height=3.0in]{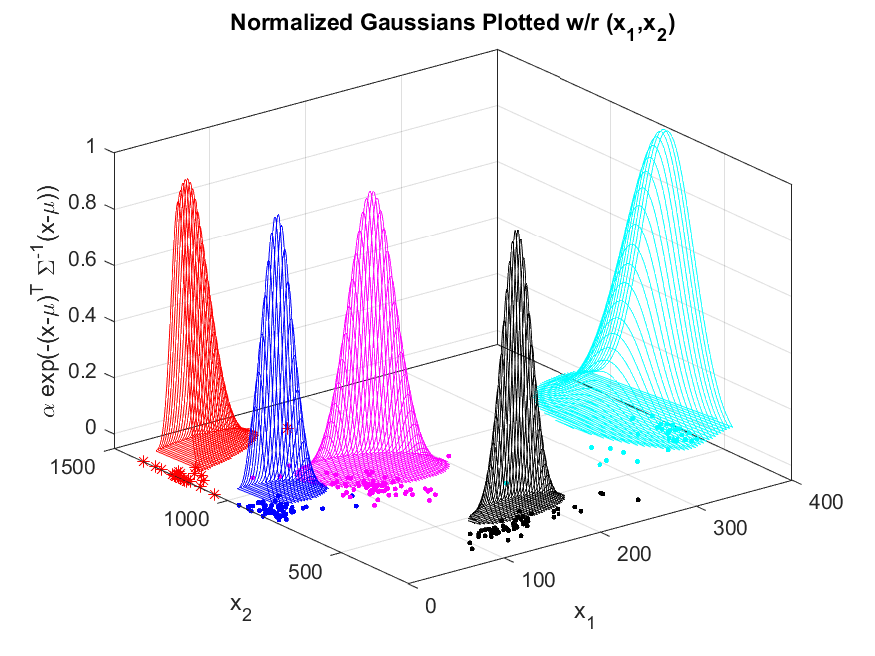}
  \caption {Data points from Figure~\ref{Data.Summary1}d
  plotted w/r to the $x_1-x_2$ axes
and their normalized Gaussian distributions. For visualization purposes, the
 data points are lowered to be just below the $x_1-x_2$ plane.
 The covariance matrices and mean values
of each formation are obtained from the geochemical data collected in the Lake Turkana region.
The green points from the {\it Has} formation are excluded because they are deemed to be too
unreliable.}
    \label{Data.Summary2}
\end{figure}

Figure~\ref{Data.Summary2}
shows the Gaussian distributions
and their associated Figure~\ref{Data.Summary1}d points in  the $x_1-x_2$ plane. It is clear that
most of the plotted points for any one formation are confined to be within 2 standard deviations
of their corresponding Gaussian distribution.
In this case, a new test point can be
classified by assigning it to the cluster that gives the highest
probability value $p(c|\x)$ in equation~\ref{GDA.eq8}.

\begin{figure}
\includegraphics[width=5.8in]{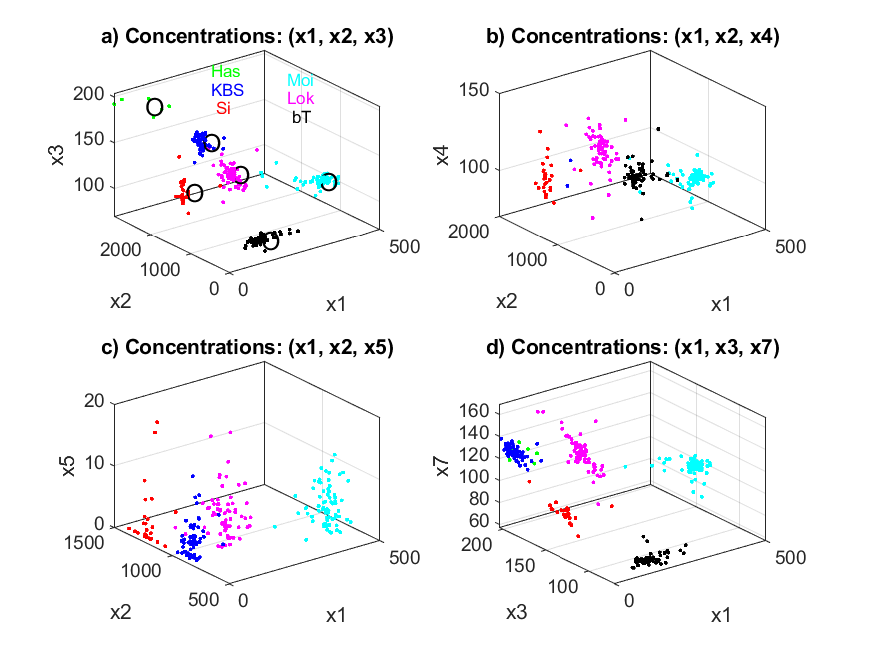}
  \caption {Same as Figure~\ref{3D.Concentration.B} except the compound indices
  $(i,j,k)\in\{(1,2,3), ~(1,2,4),~(1,3,5),~(1,3,7)\}$
  are the ones where there is no overlap between the ellipsoids
  associated with any of the formations with different colors.
  Here, $alpha=0.01$ in equation~\ref{Eq.overlap} and the Gaussian density function
is normalized to unity in equation~\ref{GDA.eq5}.}
    \label{NonOverlap.1}
\end{figure}

\begin{figure}
\includegraphics[width=5.8in]{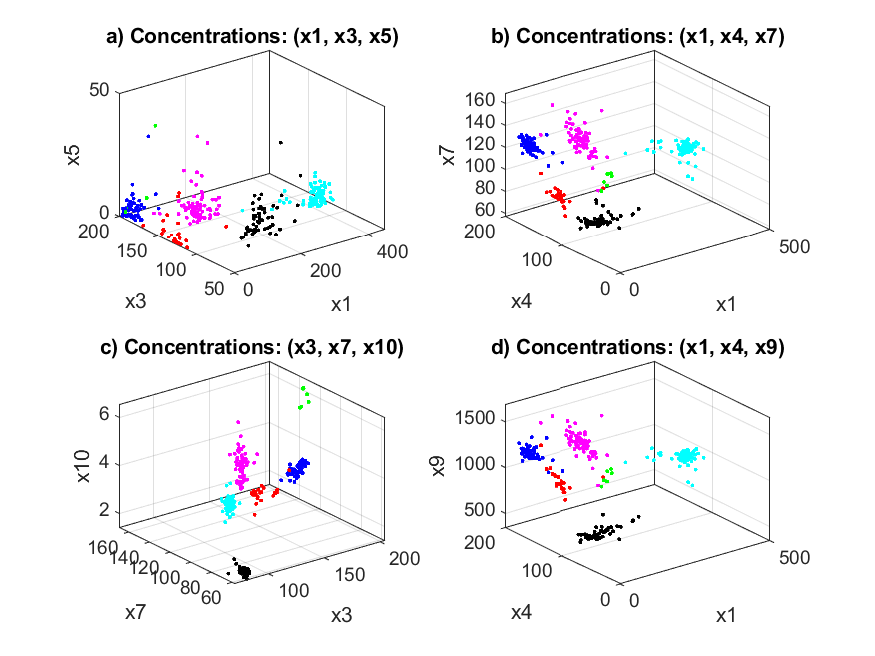}
  \caption {Same as the previous figure except
   $(i,j,k)\in\{(1,2,5), ~(1,4,7),~(3,7,10),~(1,4,9)\}$.
   The optimal coordinate system can be quantified by designing
   a metric that rewards wider separation of the
   mean vectors
   from one another as well as smaller variances in the ellipsoids
   (Subbalakshmi et al., 2015).}
    \label{NonOverlap.2}
\end{figure}

All possible triplets of compound indices $(i,j,k)$ for $i,j,k \in \{1,2,...,11\}$ were tested to identify those where the
associated Gaussian ellipsoids did not overlap.
The results are shown in Figures~\ref{NonOverlap.1} and~\ref{NonOverlap.2},
and reveal that the point clouds for any one color do not overlap those of any other color
except for the green points.
There were only 7 samples for the green points, which correspond to
 samples from the "Has" formation, so they were
excluded from this analysis.


The value of this GDA analysis is that, for the Turkana data, there are
eight triplets of coordinates, i.e. fingerprints,  that should consistently identify the same formation of new rock sample. The reliability of this identification can be quantified by
using the value of the probability function with the mean vector closest to the test sample.

\section{Summary}

We presented a search strategy that can find
unique geochemical fingerprints of formations in the Omo Group of Lake Turkana, Kenya.
Gaussian Discriminant Analysis (Bishop, 2006)  is used to fit
each geochemical point cloud to its Gaussian distribution. After fitting
we can
then efficiently identify the triplet of coordinates $(x_i,x_j,x_k)$, where
the Gaussians
don't overlap with one another within two standard deviations;
the method is not restricted to triplets of coordinates.
For the Lake Turkana geochemical data we discovered
8 sets of coordinates where the Gaussian ellipsoids
do not overlap with a probability threshold of $0.01$.
Therefore, the formation of an undated rock sample can be determined
by plotting it in one or all of the 8 independent
coordinate systems $(x_i,x_j,x_k)$.
Moreover, this procedure provides the probability value of the rock sample belonging to a formation and, unlike PCA, provides more than one coordinate
system that can be used to confirm the class of an untested data point.

If there are no non-overlapping Gaussians, then
the triplet of coordinate combinations is sought where there are only two overlapping Gaussians. In this case,
these two Gaussians for beds A and B are further tested for a  unique combination of triplet coordinates
that only allows either bed A or B, but not both.

It is important to note that while GDA can be a useful tool for identifying geochemical fingerprints of geological formations, it is not always possible to find a unique fingerprint for every formation. Factors such as mixing of materials, alteration over time, and the presence of multiple depositional environments can all complicate the geochemical signature of a formation.
Therefore, careful consideration must be given to the specific data and geological context in order to ensure accurate results. Fortunately, our procedure
tests every possible coordinate system and ignores the ones which are not informative.

Finally, our modified GDA procedure can be applied to almost any geological problem where many data samples, each associated with a high-dimensional vector, is employed for identifying the class of an unknown test sample. For example, amplitude versus offset (Yilmaz, 2001) studies plot seismic  data points in the simple intercept-slope space. This 2D space is sometimes
insufficient for accurately distinguishing an economic play from a dry hole. Therefore, multidimensional geophysical
measurements and
rock physics data such as
seismic velocity, density, porosity, conductivity, resistivity, attenuation and permeability should be used for each data point
in the depth domain of the migration image. This means that each data sample can be a 9-dimensional vector, or higher, and can be used to ascertain which 3D or 4D coordinate systems
have separable ellipsoids. Unlike a black box PCA analysis,
the selection of the separable coordinate systems can be partly guided by the interpreters
physics-informed intuition.

\section{Acknowledgements} We are very grateful to Patrick Gathogo
 who defined provided the data.
 We are also grateful to
 Prof. Bereket Haileab (bhaileab@carleton.edu) at Carleton University who
 was responsible for collecting rock samples in the Lake Turkana region and analyzing them
for geochem composition. His efforts were assisted under
the direction of Professor Frank Brown at the University of Utah.
\section{References}
\begin{itemize}

\item Bishop, C., 2006, Pattern Recognition and Machine Learning: Springer Press.

\item DeGroot, M., 1975, Probability and Statistics: Addison-Wesley Publishing Co.

\item Gathogo, P. and F. Brown, 2006,
Stratigraphy of the Koobi Fora formation (Pliocene and Pleistocene) in the Ileret region of northern Kenya:
Journal of African Earth Sciences,
{\bf 45}, 369--390.

\item  Subbalakshmi, C.,  G. Rama Krishna, S. Krishna Mohan Rao, and P. Venketeswa Rao, 2015,
A method to find optimum number of clusters based on fuzzy silhouette on dynamic data set:
Procedia Computer Science,
{ \bf 46}, 346-353.

\item   Yilmaz, {\"O}., 2001, Seismic Data Analysis: SEG Press Book.

\end{itemize}

\end{document}